\documentstyle[12pt]{article}
\begin{document}
\hfuzz=1pt
\setlength{\textheight}{8.5in}
\setlength{\topmargin}{0in}
\begin{center}
\Large {\bf Maximal violation of
Bell's inequality and atomic cascade photons 
} \\  \vspace{.75in}
\large {M. Ardehali}\footnote[1]
{email address:mardehali@iwmail.com}
\\ \vspace{.3in}
Research Laboratories,
NEC Corporation,\\
Sagamihara,
Kanagawa 229
Japan

\end{center}
\vspace{.20in}

\begin{abstract}
A correlation inequality is derived from local realism and a
supplementary assumption. 
This inequality is violated by
a factor of 1.5 
in the case of real experiments, whereas
previous inequalities such as
Clauser-Horne-Shimony-Holt inequality of $1969$
and Clauser-Horne inequality of $1974$
are violated by a factor of $\sqrt 2$.
Thus the magnitude of violation of this inequality
is approximately $20.7\%$ larger than the magnitude of violation of
previous inequalities.
Moreover, the present inequality can be used to 
test locality very simply because it requires the 
measurements of only two detection probabilities. In contrast,
Clauser-Horne inequality requires the measurements of five 
detection probabilities.
\end {abstract}
\pagebreak

\begin{center}
\Large  {\bf I. Introduction}
\end{center}

The Copenhagen interpretation of quantum mechanics
is based on the fundamental assumption that the wave
function, with its statistical interpretation, provides
a complete description of physical reality.
This assumption
has been the object of severe criticism, most notably by
Einstein, who always maintained that the wave function should
be supplemented with additional ``hidden variables'' such that
these variables together with the wave function precisely determine
the results of individual experiments. In 1965, Bell
\cite{1} showed that the premises of locality and
realism, as postulated by
Einstein, Podolsky, and Rosen (EPR) \cite{2}, imply some
constrains on the statistics of two spatially separated
particles. These constrains, which are collectively known as
Bell inequalities, are sometime grossly violated by quantum
mechanics. Bell's theorem therefore is a proof
that all realistic interpretation
of quantum mechanics must be non-local.

Bell's original argument, however,
can not be experimentally tested
because it relies on perfect correlation of the spin of the two
particles \cite {3}. Faced with this problem,
Clauser-Horne-Shimony-Holt (CHSH) \cite{4},
Freedman-Clauser (FC) \cite{5}, and Clauser-Horne (CH) \cite{6}
derived correlation
inequalities for
systems which do not achieve $100\%$ correlation,
but which do achieve a necessary minimum correlation.
Quantum mechanics
violates these inequalities
by as much as $\sqrt 2$
(for a detailed discussion of CHSH, FC and CH inequalities, see the
review article by Clauser and Shimony \cite{7}, especially
inequalities 5.3-5.7).
An experiment based on CHSH, or FC,
or CH inequality utilizes one-channel
polarizers in which the dichotomic choice is between the detection of
the photon and its lack of detection. A better experiment is
one in which
a truly binary choice
is made between the ordinary and the extraordinary rays.
In 1971, Bell \cite{8}, and later others [9-11],
derived correlation inequalities in which two-channel
polarizers
are used to test locality. Quantum mechanical probabilities
violate these
inequalities also by a factor of $\sqrt 2$.
In this paper, we derive a correlation inequality
for two-channel polarizer systems and we show that
quantum mechanics violates this inequality
by a factor of $1.5$. Thus the magnitude of
violation of the inequality derived in this paper
is approximately $20.7\%$ larger than
the magnitude of violation of previous
inequalities of [4-11]. This result can be of
importance for the experimental test of local realism.

\begin{center}
\Large  {\bf II. EPR-Bohm gedankenexperiment}
\end{center}

We start by considering Bohm's \cite{12} version of EPR experiment
in which an unstable source emits pairs of photons in
a cascade from state $J=1$ to $J=0$.
The source is viewed by two
apparatuses.
The first (second) apparatus consists of a polarizer
$P_1 \left(P_2 \right)$
set at angle $\mbox{\boldmath $a$} \left(
\mbox{\boldmath $b$} \right)$,
and two detectors
$D_{1}^{\,\pm} \left (D_{2}^{\,\pm} \right)$
put along the ordinary and the extraordinary beams.
During a period of time $T$, the source emits, say, $N$ pairs of
photons. Let $N^{\,\pm\,\pm}\left(\mbox{\boldmath $a,b$}\right)$
be the number of simultaneous counts
from detectors $D_{1}^{\pm}$ and $D_{2}^{\pm}$,
$N^{\,\pm}\left(\mbox{\boldmath $a$}\right)$
the number of counts from detectors
$D_1^\pm$, and
$N^{\,\pm}\left(\mbox{\boldmath $b$}\right)$
the number of counts from detectors
$D_2^\pm$.
If the time $T$ is sufficiently
long, then the ensemble probabilities
$p^{\;\pm\;\pm}\left(\mbox{\boldmath $a,b$}\right)$ are defined as
\begin{eqnarray}{\nonumber}
p^{\;\pm\;\pm} \left(\mbox{\boldmath $a,b$} \right)&=&
\frac{N^{\;\pm\;\pm} \left(\mbox{\boldmath $a,b$} \right)}{N}, \\ 
\nonumber
p^{\;\pm}(\mbox{\boldmath $a$})&=&
\frac{N^{\;\pm}(\mbox{\boldmath $a$})}{N}, \\ 
p^{\;\pm}(\mbox{\boldmath $b$})&=&
\frac{N^{\;\pm}(\mbox{\boldmath $b$})}{N}.
\end{eqnarray}
\noindent We consider a particular pair of photons and specify its
state with a parameter $\lambda$. Following Bell, we do not 
impose any restriction on the complexity of $\lambda$. 
``It is
a matter of indifference
in the following whether $\lambda$ denotes a single variable or
a set, or even a set of functions, and whether the variables are 
discrete or continuous \cite{1}.''

The ensemble probabilities
in Eq. $(1)$ are defined as
\begin{eqnarray} {\nonumber}
p^{\:\pm\:\pm}(\mbox{\boldmath $a,b$}) &=&
\int p\,(\lambda)\, p^{\;\pm}(\mbox{\boldmath 
$a$} \mid \lambda) \, p^{\;\pm}(\mbox{\boldmath $b$}
\mid \lambda,\mbox{\boldmath $a$}), \\ \nonumber
p^{\:\pm}(\mbox{\boldmath $a$}) &=&
\int p \, (\lambda) \, p^{\;\pm}(\mbox{\boldmath 
$a$} \mid \lambda), \\ 
p^{\:\pm}(\mbox{\boldmath $b$}) &=&
\int p \, (\lambda)\, p^{\;\pm}(\mbox{\boldmath 
$b$} \mid \lambda).
\end{eqnarray}
Equations (2) may be stated in physical terms: The ensemble
probability for detection of photons by
detectors $D^{\;\pm}_{\; 1}$ and $D^{\;\pm}_{\;2}$
[that is $p^{\;\pm\;\pm}(\mbox{\boldmath $a,b$})$]
is equal to the sum or integral of the probability
that the emission is
in the state $\lambda$ [that is $p(\lambda)$], times the conditional
probability that if the emission is in the state $\lambda$,
then a count is triggered by the first detector $D^{\;\pm}_{1}$
[that is $p^{\;\pm}(\mbox{\boldmath $a$}
\mid \lambda)$],
times the conditional probability that 
if the emission is in the state 
$\lambda$ and if the first polarizer is set along axis $\boldmath a$,
then a count is triggered from the second detector $D^{\;\pm}_{2}$
[that is $p^{\;\pm}(\mbox{\boldmath $b$}
\mid \lambda,\mbox{\boldmath $a$})$].
Similarly the ensemble probability for detection of photons by
detector $D^{\;\pm}_{\;1} \left(D^{\;\pm}_{\;2} \right )$
{\large [} that is $p^{\;\pm}(\mbox{\boldmath $a$}) \left
[p^{\;\pm}(\mbox{\boldmath $b$}) \right]$ {\large ]}
is equal to the sum or integral of the probability that the photon
is in the state $\lambda$ [that is $p(\lambda)$], times the
conditional probability that if the
photon is in the state $\lambda$,
then a count is triggered by
detector $D^{\;\pm}_{1} \left(D^{\;\pm}_{2} \right )$
{\large[} that is $p^{\;\pm}(\mbox{\boldmath $a$}
\mid \lambda) \left [p^{\;\pm}(\mbox{\boldmath $b$} \mid 
\lambda ) \right ]$ {\large]}.
Note that Eqs. $(1)$ and $(2)$ are quite general and follow
from the standard rules of probability theory.
No assumption has yet been made that is not satisfied 
by quantum mechanics.

Hereafter, we
focus our attention only on those theories that satisfy
EPR criterion of locality: `` Since at the time of measurement the
two systems no longer interact, no real change can take place in the
second system in consequence of anything that may be done to first
system
\cite {2}''. EPR's criterion of locality can be translated into
the following mathematical equation:
\begin{equation}
p^{\;\pm}(\mbox{\boldmath $b$} \mid \lambda,
\mbox{\boldmath $a$})=
p^{\;\pm}(\mbox{\boldmath $b$} \mid \lambda).
\end{equation}
Equation $(3)$ is the hall mark of local realism.
\footnote [2] {
It is worth noting that there is a
difference between Eq. (3) and CH's
criterion of locality. CH write their assumption of
locality as
\begin{eqnarray*}
p^+ \left(\mbox{\boldmath $a, b$} , \lambda \right)=
p^+ \left(\mbox{\boldmath $a$} , \lambda \right)
p^+ \left(\mbox{\boldmath $b$}, \lambda \right).
\end{eqnarray*}
Apparently by $p^+ \left(\mbox{\boldmath $a, b$} , \lambda \right)$,
they mean the conditional probability that if the
emission is in state $\lambda$,
then simultaneous counts are triggered by detectors
$D^+_1$ and $D^+_2$.  However, what they call
$p^+ \left(\mbox{\boldmath $a, b$} , \lambda \right)$
in probability theory is usually
called $p^+ \left(\mbox{\boldmath $a, b$} \mid\lambda \right)$ [note
that $p (x, y, z)$ is the joint
probability of $x, y$ and $z$, whereas $p (x, y \mid z)$
is the conditional probability that
if $z$ then $x$ and $y$]. Similarly by
$p^+ \left(\mbox{\boldmath $a$} , \lambda \right) {\large [}
p^+ \left(\mbox{\boldmath $b$}, \mid \lambda \right){\large]}$,
CH mean the conditional
probability that if the emission is in state $\lambda$,
then a count is triggered from the
detector $D^+_1 \left(D^+_2 \right)$.
Again what they call
$p^+ \left(\mbox{\boldmath $a$} , \lambda \right) {\large [}
p^+ \left(\mbox{\boldmath $b$}, \lambda \right){\large]}$
in probability
theory is usually written as
$p^+ \left(\mbox{\boldmath $a$} \mid \lambda \right) {\large [}
p^+ \left(\mbox{\boldmath $b$} \mid \lambda \right){\large]}$
(again note that $p(x, z)$ is the
joint probability of $x$ and $z$, whereas $p (x \mid z)$
is the conditional probability
that if $z$ then $x$).
Thus according to standard notation of probability theory,
CH criterion of locality may be written as
\begin{eqnarray*}
p^+ \left(\mbox{\boldmath $a, b$} \mid \lambda \right)=
p^+ \left(\mbox{\boldmath $a$} \mid \lambda \right)
p^+ \left(\mbox{\boldmath $b$} \mid \lambda \right).
\end{eqnarray*}
Now according to Bayes' theorem,
\begin{eqnarray*}
p^+ \left(\mbox{\boldmath $a, b$} \mid \lambda \right)=
p^+ \left(\mbox{\boldmath $a$} \mid \lambda \right)
p^+ \left(\mbox{\boldmath $b$} \mid \lambda,
\mbox{\boldmath $a$}, \right).
\end{eqnarray*}
Substituting the above equation in CH's criterion of locality,
we obtain
\begin{eqnarray*}
p^+ \left(\mbox{\boldmath $ b$}
\mid \lambda, \mbox{\boldmath $a$} \right)=
p^+ \left(\mbox{\boldmath $b$} \mid \lambda\right),
\end{eqnarray*}
which for the ordinary equation is the same as Eq. (3).}
It is the most general form of locality that accounts
for correlations subject only to the requirement that a count
triggered by the second detector does not depend on
the orientation of the first polarizer. The assumption
of locality, i.e., Eq. $(3)$, is
quite natural since the two photons are spatially separated so that
the orientation of the first polarizer should not influence the
measurement carried out on the second photon.

\begin{center}
\Large  {\bf III. Derivation of Bell's inequality}
\end{center}

We  now show that 
equation $(3)$ leads to validity of an equality
that is sometimes grossly violated by
the quantum mechanical predictions in the case of real experiments.
First we need to prove the following algebraic theorem.

{\it Theorem:} Given ten non-negative real numbers
$x_{1}^{+}$, $x_{1}^{-}$, $x_{2}^{+}$, $x_{2}^{-}$,
$y_{1}^{+}$, $y_{1}^{-}$, $y_{2}^{+}$, $y_{2}^{-}$, $U$ and $V$
such that
$x_{1}^{+}, x_{1}^{-},
x_{2}^{+}, x_{2}^{-} \leq U$,
and
$y_{1}^{+}, y_{1}^{-},
y_{2}^{+}, y_{2}^{-} \leq V$,
then the following inequality always holds:
\begin{eqnarray}{\nonumber}
Z &=& x_{1}^{+}y_{1}^{+}
+x_{1}^{-}y_{1}^{-}
-x_{1}^{+}y_{1}^{-}
-x_{1}^{-}y_{1}^{+}
+y_{2}^{+}x_{1}^{+}
+y_{2}^{-}x_{1}^{-} \\ \nonumber
&-&y_{2}^{+}x_{1}^{-}
-y_{2}^{-}x_{1}^{+}
+y_{1}^{+}x_{2}^{+}
+y_{1}^{-}x_{2}^{-}
-y_{1}^{+}x_{2}^{-}
-y_{1}^{-}x_{2}^{+}
-2x_{2}^{+}y_{2}^{+} \\
&-&2x_{2}^{-}y_{2}^{-}
+Vx_{2}^{+}+Vx_{2}^{-}
+Uy_{2}^{+}+Uy_{2}^{-}+UV
\ge 0.
\end{eqnarray}
{\it Proof}:
Calling $A=y_{1}^{+}-y_{1}^{-}$, we write the function $Z$ as
\begin{eqnarray} {\nonumber}
Z&=&
x_{2}^{+}
\left(-2y_{2}^{+} + A + V \right )
+ x_{2}^{-}
\left(-2y_{2}^{-} - A + V \right ) \\
&+&\left( x_{1}^{+} - x_{1}^{-} \right )
\left(A + y_{2}^{+}-y_{2}^{-} \right )
+ U y_{2}^{+} +  U y_{2}^{-}
+UV.
\end{eqnarray}

\noindent We consider the following eight cases:
\\
(1) First assume
$\left \{
\begin{array}{c}
-2y_{2}^{+} + A  + V \ge 0,\\
-2y_{2}^{-} - A + V \ge 0,\\
A + y_{2}^{+}-y_{2}^{-}\ge 0.
\end{array} \right.$
\vspace{0.4 cm}

\noindent The function $Z$ is minimized if
$x_{2}^{+}=0, x_{2}^{-}=0$, and
$ x_{1}^{+} - x_{1}^{-} =-U$. Thus
\begin{eqnarray} {\nonumber}
Z &\ge&
-U \left(A +y_{2}^{+} - y_{2}^{-} \right )
+U y_{2}^{+} +  U y_{2}^{-}
+UV \\
&=&U\left(-A + 2y_{2}^{-} + V\right ).
\end{eqnarray}
Since $V \ge A$ and $y_{2}^{-} \ge 0$, $Z \ge 0$.
\vspace{0.7 cm}
\\
(2) Next assume
$\left \{
\begin{array}{c}
-2y_{2}^{+} + A + V < 0,\\
-2y_{2}^{-} - A + V \ge 0,\\
A + y_{2}^{+}-y_{2}^{-}\ge 0.
\end{array} \right.$
\vspace{0.4 cm}

\noindent The function $Z$ is minimized if
$x_{2}^{+}=U, x_{2}^{-}=0$, and
$ x_{1}^{+} - x_{1}^{-} =-U$. Thus
\begin{eqnarray} {\nonumber}
Z &\ge&
U
\left(-2y_{2}^{+} + A + V \right ) -
U \left(A + y_{2}^{+}-y_{2}^{-} \right )
+ U y_{2}^{+} +  U y_{2}^{-}
+UV \\
&=&2U\left(V+y_{2}^{-}-y_{2}^{+}\right ).
\end{eqnarray}
Since $V \ge y_{2}^{+}$, and $y_{2}^{-} \ge 0$, $Z \ge 0$.
\vspace{0.7 cm}
\\
(3) Next assume
$\left \{
\begin{array}{c}
-2y_{2}^{+} + A + V \ge 0,\\
-2y_{2}^{-} - A + V < 0,\\
A + y_{2}^{+}-y_{2}^{-}\ge 0.
\end{array} \right.$
\vspace{0.4 cm}

\noindent The function $Z$ is minimized if
$x_{2}^{+}=0, x_{2}^{-}=U$, and
$ x_{1}^{+} - x_{1}^{-} =-U$. Thus
\begin{eqnarray} {\nonumber}
Z &\ge&
U
\left(-2y_{2}^{-} - A + V \right ) -
U \left(A + y_{2}^{+}-y_{2}^{-} \right )
+ U y_{2}^{+} +  U y_{2}^{-}
+UV \\
&=&2U\left (V - A \right).
\end{eqnarray}
Since $V \ge  A$, $Z \ge 0$.
\vspace{0.7 cm}
\\
(4) Next assume
$\left \{
\begin{array}{c}
-2y_{2}^{+} + A + V \ge 0,\\
-2y_{2}^{-} - A + V \ge 0,\\
A + y_{2}^{+}-y_{2}^{-} < 0.
\end{array} \right.$
\vspace{0.4 cm}

\noindent The function $Z$ is minimized if
$x_{2}^{+}=0, x_{2}^{-}=0$, and
$ x_{1}^{+} - x_{1}^{-} =U$. Thus
\begin{eqnarray} {\nonumber}
Z &\ge&
U \left(A + y_{2}^{+}-y_{2}^{-} \right )
+ U y_{2}^{+} +  U y_{2}^{-}
+UV \\
&=&U\left(A + 2y_{2}^{+} + V\right ).
\end{eqnarray}
Since $V \ge A$ and $y_{2}^{+} \ge 0$, $Z \ge 0$.
\vspace{0.7 cm}
\\
(5) Next assume
$\left \{
\begin{array}{c}
-2y_{2}^{+} + A + V < 0,\\
-2y_{2}^{-} - A + V < 0,\\
A + y_{2}^{+}-y_{2}^{-}\ge 0.
\end{array} \right.$
\vspace{0.4 cm}

\noindent The function $Z$ is minimized if
$x_{2}^{+}=U, x_{2}^{-}=U$, and
$ x_{1}^{+} - x_{1}^{-} =-U$. Thus
\begin{eqnarray} {\nonumber}
Z &\ge&
U
\left(-2y_{2}^{+} + A + V \right )
+U
\left(-2y_{2}^{-} - A + V \right )
-U \left(A + y_{2}^{+}-y_{2}^{-} \right )\\ \nonumber
&+& U y_{2}^{+} +  U y_{2}^{-}
+UV \\ 
&=&U\left(-2y_{2}^{+} -A +3V \right ).
\end{eqnarray}
Since $V \ge A$ and $V \ge y_{2}^{+}$, $Z \ge 0$.
\vspace{0.7 cm}
\\
(6) Next assume
$\left \{
\begin{array}{c}
-2y_{2}^{+} + A + V < 0,\\
-2y_{2}^{-} - A + V \ge 0,\\
A + y_{2}^{+}-y_{2}^{-} < 0.
\end{array} \right.$
\vspace{0.4 cm}

\noindent The function $Z$ is minimized if
$x_{2}^{+}=U, x_{2}^{-}=0$, and
$ x_{1}^{+} - x_{1}^{-} =U$. Thus
\begin{eqnarray} {\nonumber}
Z &\ge&
U\left(-2y_{2}^{+} + A + V \right )+
U \left(A + y_{2}^{+}-y_{2}^{-} \right )
+ U y_{2}^{+} +  U y_{2}^{-}
+UV \\ 
&=&2U\left(A + V \right ).
\end{eqnarray}
Since $V \ge A$, $Z \ge 0$.
\vspace{0.7 cm}
\\
(7) Next assume
$\left \{
\begin{array}{c}
-2y_{2}^{+} + A + V \ge 0,\\
-2y_{2}^{-} - A + V < 0,\\
A + y_{2}^{+}-y_{2}^{-} < 0.
\end{array} \right.$
\vspace{0.4 cm}

\noindent The function $Z$ is minimized if
$x_{2}^{+}=0, x_{2}^{-}=U$, and
$ x_{1}^{+} - x_{1}^{-} = U$. Thus
\begin{eqnarray} {\nonumber}
Z &\ge&
U
\left(-2y_{2}^{-} - A + V \right )
+U \left(A + y_{2}^{+}-y_{2}^{-} \right )
+ U y_{2}^{+} +  U y_{2}^{-}
+UV \\ 
&=&2U\left( y_{2}^{+}-y_{2}^{-} + V \right ).
\end{eqnarray}
Since $V \ge y_{2}^{-}$ and $y_{2}^{+} \ge 0$, $Z \ge 0$.
\vspace{0.7 cm}
\\
(8) Finally assume
$\left \{
\begin{array}{c}
-2y_{2}^{+} + A + V < 0,\\
-2y_{2}^{-} - A + V < 0,\\
A + y_{2}^{+}-y_{2}^{-} < 0.
\end{array} \right.$
\vspace{0.3 cm}

\noindent The function $Z$ is minimized if
$x_{2}^{+}=U, x_{2}^{-}=U$, and
$ x_{1}^{+} - x_{1}^{-} =U$. Thus
\begin{eqnarray} {\nonumber}
Z &\ge&
U
\left(-2y_{2}^{+} + A +V \right )
+U
\left(-2y_{2}^{-} - A + V \right )
+U
\left(A + y_{2}^{+}-y_{2}^{-} \right ) \\ \nonumber
&+& U y_{2}^{+} +  U y_{2}^{-}
+UV \\
&=& U\left(-2y_{2}^{-} + A + 3V \right ).
\end{eqnarray}
Since $V \ge A$ and $V \ge y_{2}^{-}$, $Z \ge 0$,
and the theorem is proved.

Now let $\mbox {\boldmath $a$ ($b$)}$
and $\mbox {\boldmath $a'$ ($b'$)}$
be two arbitrary orientation of the first
(second) polarizer, and let
\begin{eqnarray}{\nonumber}
x_{1}^{\pm}&=&p^{\;\pm}(\mbox{\boldmath $a$} \mid \lambda), \qquad
x_{2}^{\pm}=p^{\;\pm}(\mbox{\boldmath $a'$}|\lambda), \\ 
y_{1}^{\pm}&=&p^{\;\pm}(\mbox{\boldmath $b$}|\lambda), \qquad
y_{2}^{\pm}=p^{\;\pm}(\mbox{\boldmath $b'$}|\lambda).
\end{eqnarray}

\noindent Obviously for each value of $\lambda$, we have
\begin{eqnarray}{\nonumber}
p^{\;\pm}(\mbox{\boldmath $a$} \mid \lambda) \leq 1, \qquad
p^{\;\pm}(\mbox{\boldmath $a'$} \mid \lambda) \leq 1,\\ 
p^{\;\pm}(\mbox{\boldmath $b$} \mid \lambda) \leq 1, \qquad
p^{\;\pm}(\mbox{\boldmath $b'$} \mid \lambda) \leq 1.
\end{eqnarray}

\noindent Inequalities ($4$) and ($15$) yield
\begin{eqnarray} {\nonumber}
&&p^{+}(\mbox{\boldmath $a$} \mid \lambda) \,
p^{+}(\mbox{\boldmath $b$} \mid \lambda)
+p^{-}(\mbox{\boldmath $a$} \mid \lambda) \,
p^{-}(\mbox{\boldmath $b$} \mid \lambda)
-p^{+}(\mbox{\boldmath $a$} \mid \lambda) \,
p^{-}(\mbox{\boldmath $b$} \mid \lambda)  \\ \nonumber
&&- \,p^{-}(\mbox{\boldmath $a$} \mid \lambda) \, 
p^{+}(\mbox{\boldmath $b$} \mid \lambda)
+p^{+}(\mbox{\boldmath $b'$} \mid \lambda) \,
p^{+}(\mbox{\boldmath $a$} \mid \lambda) 
+p^{-}(\mbox{\boldmath $b'$} \mid \lambda) \,
p^{-}(\mbox{\boldmath $a$} \mid \lambda) \\ \nonumber
&&- \, p^{+}(\mbox{\boldmath $b'$} \mid \lambda) \,
p^{-}(\mbox{\boldmath $a$} \mid \lambda) -
p^{-}(\mbox{\boldmath $b'$} \mid \lambda) \,
p^{+}(\mbox{\boldmath $a$} \mid \lambda) +
p^{+}(\mbox{\boldmath $b$} \mid \lambda) \,
p^{+}(\mbox{\boldmath $a'$} \mid \lambda) \\ \nonumber
&&+ \, p^{-}(\mbox{\boldmath $b$} \mid \lambda) \,
p^{-}(\mbox{\boldmath $a'$} \mid \lambda) -
p^{+}(\mbox{\boldmath $b$} \mid \lambda) \,
p^{-}(\mbox{\boldmath $a'$} \mid \lambda) 
-p^{-}(\mbox{\boldmath $b$} \mid \lambda) \,
p^{+}(\mbox{\boldmath $a'$} \mid \lambda) \\ \nonumber
&&- \, 2p^{+}(\mbox{\boldmath $a'$} \mid \lambda) \, 
p^{+}(\mbox{\boldmath $b'$} \mid \lambda)-
2p^{-}(\mbox{\boldmath $a'$} \mid \lambda) \,
p^{-}(\mbox{\boldmath $b'$} \mid \lambda)+ 
p^{+}(\mbox{\boldmath $a'$} \mid \lambda) \\
&&+ \, p^{-}(\mbox{\boldmath $a'$} \mid \lambda)
+p^{+}(\mbox{\boldmath $b'$} \mid \lambda) \,
+p^{-}(\mbox{\boldmath $b'$} \mid \lambda) \ge -1.
\end{eqnarray}

\noindent Multiplying both sides of $(16)$
by $p \, (\lambda)$, integrating over $\lambda$ and
using Eqs. $(2)$, we obtain
\begin{eqnarray} {\nonumber}
&&p^{+ +}(\mbox{\boldmath $a,\, b$}) 
+p^{- \,-}(\mbox{\boldmath $a,\, b$}) 
-p^{+ \,-}(\mbox{\boldmath $a,\, b$}) 
-p^{- \,+}(\mbox{\boldmath $a,\, b$})  
+p^{+ +}(\mbox{\boldmath $b',\, a$})+ \\ \nonumber
&&p^{- \,-}(\mbox{\boldmath $b',\, a$})
-p^{+ \,-}(\mbox{\boldmath $b',\, a$})
-p^{- \,+}(\mbox{\boldmath $b',\, a$}) 
+p^{+ +}(\mbox{\boldmath $b,\, a'$})+ \\ \nonumber
&&p^{- \,-}(\mbox{\boldmath $b,\, a'$}) 
-p^{+ \,-}(\mbox{\boldmath $b,\, a'$}) 
-p^{- \,+}(\mbox{\boldmath $b,\, a'$}) 
-2p^{+ +}(\mbox{\boldmath $a',\, b'$}) - \\
&&2p^{- \,-}(\mbox{\boldmath $a',\, b'$}) 
+p^{+}(\mbox{\boldmath $a'$}) 
+p^{-}(\mbox{\boldmath $a'$}) 
+p^{+}(\mbox{\boldmath $b'$}) 
+p^{-}(\mbox{\boldmath $b'$})
\geq -1.
\end{eqnarray}
We now note that 
the expected value of detection probabilities while polarizers
are set along orientations $\mbox{\boldmath $m$}$ and
$\mbox{\boldmath $n$}$, i.e., 
$E \,\left (\mbox{\boldmath $m,n$} \right)$
is defined as
\begin{eqnarray} \nonumber
E \,\left (\mbox{\boldmath $m,n$} \right) &= &
p^{+\, +} \,\left (\mbox{\boldmath $m,n$} \right)-
p^{+\, -} \,\left (\mbox{\boldmath $m,n$} \right) \\
&-&p^{-\, +} \,\left (\mbox{\boldmath $m,n$} \right)+
p^{-\, -} \,\left (\mbox{\boldmath $m,n$} \right).
\end{eqnarray}
Using $(18)$, 
inequality $(17)$ may be written as
\begin{eqnarray} {\nonumber}
&&E(\mbox{\boldmath $a,\, b$}) +
E(\mbox{\boldmath $b',\, a$})
+E(\mbox{\boldmath $b,\, a'$}) 
-2p^{+ +}(\mbox{\boldmath $a',\, b'$}) - 
2p^{- \,-}(\mbox{\boldmath $a',\, b'$}) \\
&&+p^{+}(\mbox{\boldmath $a'$}) 
+p^{-}(\mbox{\boldmath $a'$}) 
+p^{+}(\mbox{\boldmath $b'$}) 
+p^{-}(\mbox{\boldmath $b'$})
\geq -1.
\end{eqnarray} 
All local realistic theories must satisfy inequality $(17)$ [or $(19)$].

\begin{center}
\Large  {\bf IV.
Violation of Bell's inequality in case of ideal experiments}
\end{center}

First we consider an 
atomic cascade experiment in which
polarizers and detectors are ideal. 
Assuming
polarizers are set along axes $\mbox{\boldmath $m$}$
and $\mbox{\boldmath $n$}$ where $\theta=
\mid \mbox{\boldmath $m$}-\mbox{\boldmath $n$} \mid$,
the expected values, 
the single and joint detection
probabilities for a pair of photons in
a cascade from state $J=1$ to $J=0$
are given by
\begin{eqnarray} {\nonumber}
&&E \left ( \mbox{\boldmath $m,\, n$} \right)=
E \left(\theta \right) = \cos 2 \theta, \qquad
p^{+} (\mbox{\boldmath $a'$})=
p^{-} (\mbox{\boldmath $a'$})=
p^{+} (\mbox{\boldmath $b'$})=
p^{-} (\mbox{\boldmath $b'$})= \frac{1}{2}, \\ \nonumber
&&p^{+\, +} \left ( \mbox{\boldmath $m,\, n$} \right)=
p^{+\, +}\left(\theta \right) =
\frac{\cos^2 \theta}{2},
\qquad
p^{-\, -} \left ( \mbox{\boldmath $m,\, n$} \right)=
p^{-\, -} \left(\theta \right)
=\frac{\cos^2 \theta}{2}.
\\
\end{eqnarray}
Now if we choose the following orientation
$\left ( \mbox{\boldmath $a , \, b$} \right)=
\left( \mbox{\boldmath $b' , \, a$} \right) =
\left( \mbox{\boldmath $b , \, a'$}\right) = 120^\circ $
and
$\left( \mbox {\boldmath $a', \, b'$} \right) =0^\circ $
inequality $(19)$ 
becomes
\begin{eqnarray} \nonumber
&&3 E \left (120^\circ \right) 
-2 p^{+\, +} \left (0^\circ \right) 
-2 p^{-\, -} \left (0^\circ \right)+\\ 
&&p^{+} (\mbox{\boldmath $a'$})+
p^{-} (\mbox{\boldmath $a'$})+
p^{+} (\mbox{\boldmath $b'$})+
p^{-} (\mbox{\boldmath $b'$}) \ge -1.
\end{eqnarray}
Using $(20)$, we obtain
\begin{eqnarray} {\nonumber}
&&3 \cos \left (240^\circ \right)
 -2 \frac{\cos^2 \left (0^\circ \right)}
{2}
-2 \frac{\cos^2 \left (0^\circ \right)}{2} + \frac{1}{2}
 + \frac{1}{2}
 + \frac{1}{2}
 + \frac{1}{2}\\ 
&&=3*(-0.5)-2*\frac{1}{2}-2*\frac{1}{2}+2 \ge -1
\end{eqnarray}
or
\begin{eqnarray} 
-1.5 \ge -1
\end{eqnarray} 
which violates inequality $(19)$
by a factor of $1.5$
in the case of ideal experiments.

It is worth noting that
for ideal polarizers and detectors,
inequality $(19)$ is
equivalent to CHSH inequality.
In an ideal experiment, all emitted photons are analyzed
by the detectors
and hence the probability that a
photon is not collected is zero, i.e.,
\begin{eqnarray} 
p^{\pm\, 0} \,\left ( \mbox{\boldmath $m, n$} \right)=
p^{0\, \pm} \,\left ( \mbox{\boldmath $m, n$} \right)=
p^{0\, 0} \,\left ( \mbox{\boldmath $m, n$} \right)=0,
\end{eqnarray}
where for example 
$p^{\pm\, 0} \,\left ( \mbox{\boldmath $m, n$} \right)$
{\large [} $p^{0\, \pm}
\,\left ( \mbox{\boldmath $m, n$} \right)$
{\large]}
is the probability that
a count is triggered by the first (second) detector $D^{\;\pm}_{1}$
($D^{\;\pm}_{2}$)
and no count is triggered by the second (first) detector $D^{\;\pm}_{2}$
($D^{\;\pm}_{1}$),
and $p^{0 \, 0}
\,\left ( \mbox{\boldmath $m, n$} \right)$
 is the probability that no count is
triggered by the detectors.
Thus in an ideal experiment,
\begin{eqnarray} \nonumber
&&p^{+} \,\left (\mbox{\boldmath $a'$} \right)=
p^{+\, +} \,\left ( \mbox{\boldmath $a', \, b'$} \right)+
p^{+\, -} \,\left ( \mbox{\boldmath $a', \, b'$} \right),
\\
&&p^{-} \,\left (\mbox{\boldmath $a'$} \right)=
p^{-\, +} \,\left ( \mbox{\boldmath $a', \, b'$} \right)+
p^{-\, -} \,\left ( \mbox{\boldmath $a', \, b'$} \right).
\end{eqnarray}
Substituting $(25)$ in $(19)$, we obtain
\begin{eqnarray} {\nonumber}
&&E(\mbox{\boldmath $a,\, b$}) +
E(\mbox{\boldmath $b',\, a$})
+E(\mbox{\boldmath $b,\, a'$})-
p^{+\, +}(\mbox{\boldmath $a',\, b'$}) +
p^{+\, -}(\mbox{\boldmath $a',\, b'$}) + \\
&&p^{- \,+}(\mbox{\boldmath $a',\, b'$})-
p^{- \,-}(\mbox{\boldmath $a',\, b'$})
\geq -1 -p^{+} \,\left (\mbox{\boldmath $b'$} \right)
-p^{-} \,\left (\mbox{\boldmath $b'$} \right).
\end{eqnarray}
Since we have assumed detectors and polarizers are ideal, we have
$-p^{+} \,\left (\mbox{\boldmath $b'$} \right)
-p^{-} \,\left (\mbox{\boldmath $b'$} \right)=-1$. Thus
\begin{eqnarray}
E(\mbox{\boldmath $a,\, b$}) +
E(\mbox{\boldmath $b',\, a$})
+E(\mbox{\boldmath $b,\, a'$})-
E(\mbox{\boldmath $a',\, b'$})
\geq -2,
\end{eqnarray}
which are the same as CHSH inequality.

It is important to emphasize that in the case of ideal experiments,
neither the present
inequality nor CHSH are necessary: they both immediately
reduce to 
Bell's original inequality of 1965 \cite{1}. 
To show this
we assume 
$\mbox{\boldmath $a'$}$ and $\mbox{\boldmath $b'$}$ 
are along the same direction.
Since we have assumed polarizers and detectors are ideal, Eqs $(20)$
holds, and we have
$E(\mbox{\boldmath $a',\, b'$})=1$.
CHSH inequality $(27)$ therefore
becomes
\begin{eqnarray}
E(\mbox{\boldmath $a,\, b$})+
E(\mbox{\boldmath $b',\, a$})+
E(\mbox{\boldmath $b,\, a'$})
\ge -1.
\end{eqnarray}
which is the same as Bell's inequality of $1965$ \cite{1}.
Similar argument shows that for an ideal experiment,
inequality $(19)$ reduces to Bell's original inequality \cite{1}.

We have therefore shown that for
ideal polarizers and
detectors where the polarizations of two photons along
the same axis are perfectly correlated,
Bell's original inequality \cite{1} is sufficient and
there is no need for inequality $(19)$ or CHSH inequality.
Moreover, we have 
shown that for the case ideal experiments (see $24$), i. e., 
for the case where
$p^{+\, +}(\mbox{\boldmath $m,\, n$})+
p^{+\, -}(\mbox{\boldmath $m,\, n$})+
p^{-\, +}(\mbox{\boldmath $m,\, n$})+
p^{+\, -}(\mbox{\boldmath $m,\, n$}) =1$,
inequality $(19)$ is equivalent to CHSH inequality.
However, for the case of
real experiments where 
$p^{\pm\, 0}(\mbox{\boldmath $m,\, n$})$,
$p^{0\, \pm}(\mbox{\boldmath $m,\, n$})$, and
$p^{0\, 0}(\mbox{\boldmath $m,\, n$})$ are non-zero
, i. e., for the case where
$p^{+\, +}(\mbox{\boldmath $m,\, n$})+
p^{+\, -}(\mbox{\boldmath $m,\, n$})+
p^{-\, +}(\mbox{\boldmath $m,\, n$})+
p^{+\, -}(\mbox{\boldmath $m,\, n$}) <1$,
inequality $(19)$  is a distinct and new inequality
and is not equivalent to CHSH correlation.

\begin{center}
\Large  {\bf V. Necessity of a supplementary assumption for
violation of Bell's inequality in case of real experiments}
\end{center}

We now consider the violation of Bell's inequality $(19)$
in the case of real experiments, i.e., in the case of real 
detectors and polarizers.
We consider an atomic cascade experiment in which
an atom emits two photons in 
a cascade from state $J=1$ to $J=0$. Since the pair of photons
have zero angular momentum, they propagate in the form of spherical
wave. Thus the probability $p \left(\mbox{\boldmath $d_1$},
\mbox{\boldmath $d_2$} \right)$ 
of both photons being simultaneously detected
by two detectors in the directions $\mbox{\boldmath $d_1$}$ and
$\mbox{\boldmath $d_2$}$ is  \cite{3},\cite{4}
\begin{eqnarray}
p \left(\mbox{\boldmath $d_1,\,d_2$} \right)=
\eta^2 \left ({\frac{\displaystyle \Omega}
{\displaystyle 4\pi}}\right) ^2
g \left (\theta,\phi \right ),
\end{eqnarray}
where $\eta$ is the quantum efficiency of the detectors, 
$\Omega$ is the solid angle of the detector, 
$\cos \theta=\mbox{\boldmath $d_1. d_1$}$,
and angle $\phi$ is related to $\Omega$ by
\begin{eqnarray}
\Omega=2 \pi \left (1-\cos \phi \right).
\end{eqnarray}
Finally the function 
$g \left (\theta,\phi \right )$ is the angular correlation function
and in the special case is given by \cite{4}
\begin{eqnarray} 
g \left (\pi, \phi \right ) = 1+
\frac{1}{8} \cos^2 \phi \left (1 + \cos \phi \right)^2.
\end{eqnarray}
If we insert polarizers in front of the detectors, then the
quantum mechanical predictions for single and
joint detection probabilities are \cite{3},
\cite{4}
\begin{eqnarray} {\nonumber}
p^{+} \left ( \mbox{\boldmath $a$} \right )=
p^{-} \left ( \mbox{\boldmath $a$} \right )=
\eta \left ({\frac{\displaystyle \Omega}{\displaystyle 8 \pi}}
\right), \qquad
p^{+} \left ( \mbox{\boldmath $b$} \right )=
p^{-} \left ( \mbox{\boldmath $b$} \right )=
\eta \left ({\frac{\displaystyle \Omega}{\displaystyle 8 \pi}}
\right), \\ \nonumber
p^{+ \, +} \left ( \mbox{\boldmath $a,\, b$} \right )=
p^{- \, -} \left ( \mbox{\boldmath $a,\, b$} \right )=
\eta^2 \left ({\frac{\displaystyle \Omega}{\displaystyle 8 \pi}}
\right)^2
g \left (\theta,\phi \right )
\left[1+F \left (\theta,\phi \right )
\cos 2 \left ( \mbox{\boldmath $a- b$} \right ) \right ], \\  \nonumber
p^{+ \, -} \left ( \mbox{\boldmath $a,\, b$} \right )=
p^{- \, +} \left ( \mbox{\boldmath $a,\, b$} \right )=
\eta^2 \left ({\frac{\displaystyle \Omega}
{\displaystyle 8 \pi}}\right)^2
g \left (\theta,\phi \right )
\left[1-F \left (\theta,\phi \right )
\cos 2\left ( \mbox{\boldmath $a- b$} \right ) \right],
\\
\end {eqnarray}
where $F \left (\theta,\phi \right )$ is the so-called depolarization
factor and for the special case $\theta=\pi$ and small $\phi$
is given by
\begin{eqnarray}
F (\pi, \phi) \approx 1- \frac{2}{3} \left (1-\cos \phi
\right)^{2}.
\end{eqnarray}
The function $F \left (\theta,\phi \right )$
, in general, is very close to $1$ (the detailed expression for
$F \left (\theta,\phi \right )$ is given in \cite {4}).

In the atomic cascade
experiments which are feasible with present technology [5,13],
because
$\frac{\displaystyle \Omega}{\displaystyle  4 \pi} \ll 1$,
only a very small fraction of photons are detected.
Thus inequality
$(17)$ or $(19)$ can not be used to test the violation of Bell's 
inequality. 
It is important to emphasize that
a supplementary assumption is required primarily
because the solid angle covered by the
aperture of the apparatus,
$\Omega$, is  much less than $4 \pi$ and not because the
efficiency of the detectors, $\eta$, is much smaller than $1$. In
fact in the previous experiments (Ref. $13$),
the efficiency of detectors were larger than
$90 \%$.
However, because
$\frac{\displaystyle \Omega}{\displaystyle  4 \pi} \ll 1$,
all previous experiments needed
supplementary assumptions to test locality.

It is worth nothing that
CHSH \cite{4} and CH \cite{6} combined the solid angle covered
by the aperture of the apparatus 
$\frac{\displaystyle \Omega}{\displaystyle  4 \pi}$ and the
efficiency of the detectors $\eta$ into one term and wrote
$\eta_{CHSH}=\eta \frac 
{\displaystyle \Omega}{\displaystyle 4 \pi}$. They then referred to
$\eta_{CHSH}$ as efficiency of the detectors (this terminology 
however, is not usually used in optics. In optics, the efficiency of 
detector refers to the probability of
detection of a photon; it does not refer to the product of 
the solid angle covered by the detector and the
probability of detection of a photon).
CHSH and CH then pointed that
since $\eta_{CHSH} \ll 1$,
a supplementary assumption is required. To clarify
CHSH and CH argument, it should be emphasized that a supplementary
assumption is needed mainly because 
$\frac{\displaystyle \Omega}{\displaystyle  4 \pi} \ll 1$, not
because $\eta \ll 1$.

We now
state a supplementary assumption and we
show that this assumption is sufficient to make these
experiments (where $\frac{\displaystyle \Omega}
{\displaystyle  4 \pi} \ll 1$)
applicable as a test of local theories.
The supplementary assumption is:
For every emission $\lambda$, the detection probability 
by detector $D^{+}$ (or $D^-$) 
is {\it less than or equal} to the sum  of detection probabilities
by detectors $D^{+}$ and $D^-$ when the polarizer
is set along any {\it arbitrary} axis.
If we let $\mbox{\boldmath $r$}$ be an an {\it arbitrary} direction
of the 
first or second polarizer,
then the above supplementary assumption
may be translated into the 
following inequalities
\begin{eqnarray} {\nonumber}
p^{\;+}(\mbox{\boldmath $a$} \mid \lambda) \leq 
p^{\;+}(\mbox{\boldmath $r$} \mid \lambda)+
p^{\;-}(\mbox{\boldmath $r$} \mid \lambda), \qquad
p^{\;-}(\mbox{\boldmath $a$} \mid \lambda) \leq 
p^{\;+}(\mbox{\boldmath $r$} \mid \lambda)+
p^{\;-}(\mbox{\boldmath $r$} \mid \lambda), \\ \nonumber
p^{\;+}(\mbox{\boldmath $a'$} \mid \lambda) \leq 
p^{\;+}(\mbox{\boldmath $r$} \mid \lambda)+
p^{\;-}(\mbox{\boldmath $r$} \mid \lambda), \qquad
p^{\;-}(\mbox{\boldmath $a'$} \mid \lambda) \leq 
p^{\;+}(\mbox{\boldmath $r$} \mid \lambda)+
p^{\;-}(\mbox{\boldmath $r$} \mid \lambda), \\ \nonumber
p^{\;+}(\mbox{\boldmath $b$} \mid \lambda) \leq 
p^{\;+}(\mbox{\boldmath $r$} \mid \lambda)+
p^{\;-}(\mbox{\boldmath $r$} \mid \lambda), \qquad
p^{\;-}(\mbox{\boldmath $b$} \mid \lambda) \leq 
p^{\;+}(\mbox{\boldmath $r$} \mid \lambda)+
p^{\;-}(\mbox{\boldmath $r$} \mid \lambda), \\  \nonumber
p^{\;+}(\mbox{\boldmath $b'$} \mid \lambda) \leq 
p^{\;+}(\mbox{\boldmath $r$} \mid \lambda)+
p^{\;-}(\mbox{\boldmath $r$} \mid \lambda), \qquad
p^{\;-}(\mbox{\boldmath $b'$} \mid \lambda) \leq 
p^{\;+}(\mbox{\boldmath $r$} \mid \lambda)+
p^{\;-}(\mbox{\boldmath $r$} \mid \lambda).
\\
\end{eqnarray}
This supplementary assumption is obviously valid for an ensemble
of photons. 
The sum of detection probability
by detector $D^{+}$ and $D^-$
for an ensemble of photons
when the polarizer
is set along any {\it arbitrary} axis $\mbox{\boldmath $v$}$
\begin{eqnarray}
p^{+}(\mbox{\boldmath $v$})+p^{-}(\mbox{\boldmath $v$})=
\eta \left ({\frac{\displaystyle \Omega}{\displaystyle 4 \pi}} \right),
\end{eqnarray} 
whereas 
\begin{eqnarray} 
p^{+}(\mbox{\boldmath $v$})=p^{-}(\mbox{\boldmath $v$})=
\eta \left ({\frac{\displaystyle \Omega}{\displaystyle 8 \pi}} \right).
\end{eqnarray} 
The supplementary assumption requires that the corresponding
probabilities
be valid for each $\lambda$.

It is worth noting that the present supplementary assumption is
weaker than Garuccio-Rapisarda (GR) \cite{9} assumption,
that is, an experiment based on the present supplementary assumption
refutes a larger family of hidden variable theories than an 
experiment based on GR assumption.
The GR assumption is
\begin{eqnarray}
p^{+}(\mbox{\boldmath $a$} \mid \lambda)+
p^{-}(\mbox{\boldmath $a$} \mid \lambda)=
p^{+}(\mbox{\boldmath $r$} \mid \lambda)+
p^{-}(\mbox{\boldmath $r$} \mid \lambda)
\end{eqnarray}
It is easy to show that
GR assumption implies the
assumption of this paper. First note that
\begin{eqnarray} \nonumber
p^{+}(\mbox{\boldmath $a$} \mid \lambda) \le
p^{+}(\mbox{\boldmath $a$} \mid \lambda) +
p^{-}(\mbox{\boldmath $a$} \mid \lambda), \qquad
p^{-}(\mbox{\boldmath $a$} \mid \lambda) \le
p^{+}(\mbox{\boldmath $a$} \mid \lambda) +
p^{-}(\mbox{\boldmath $a$} \mid \lambda). \\
\end{eqnarray}
Now using GR assumption $(37)$,
we can immediately conclude that
\begin{eqnarray} \nonumber
p^{+}(\mbox{\boldmath $a$} \mid \lambda) \le
p^{+}(\mbox{\boldmath $r$} \mid \lambda)+
p^{-}(\mbox{\boldmath $r$} \mid \lambda), \qquad
p^{-}(\mbox{\boldmath $a$} \mid \lambda) \le
p^{+}(\mbox{\boldmath $r$} \mid \lambda)+
p^{-}(\mbox{\boldmath $r$} \mid \lambda), \\
\end{eqnarray}
which is the same as $(34)$.
Thus an
experiment which refutes the
hidden variable theories which are consistent with GR assumption also
refutes the hidden variable theories which are consistent with the
present assumption. The reverse however is not true. An experiment
based on the present supplementary assumption refutes a larger
family of hidden variable theories than an experiment based on GR
assumption. 

Now using relations $(4)$, $(14)$ and $(34)$, and applying
the same argument that
led to inequality $(17)$, we obtain the following inequality
\begin{eqnarray} {\nonumber}
&&\bigg[p^{+ \,+}(\mbox{\boldmath $a,\, b$})
+p^{- \,-}(\mbox{\boldmath $a,\, b$})
-p^{+ \,-}(\mbox{\boldmath $a,\, b$})
-p^{- \,+}(\mbox{\boldmath $a,\, b$}) 
+p^{+ +}(\mbox{\boldmath $b',\, a$})   
+p^{- \,-}(\mbox{\boldmath $b',\, a$})   \\  \nonumber
&&-p^{+ \,-}(\mbox{\boldmath $b',\, a$}) 
-p^{- \,+}(\mbox{\boldmath $b',\, a$}) 
+p^{+ \,+}(\mbox{\boldmath $b,\, a'$}) 
+p^{- \,-}(\mbox{\boldmath $b,\, a'$})
-p^{+ \,-}(\mbox{\boldmath $b,\, a'$}) \\  \nonumber
&&-p^{- \,+}(\mbox{\boldmath $b,\, a'$}) 
-2p^{+ +}(\mbox{\boldmath $a',\, b'$})
-2p^{- \,-}(\mbox{\boldmath $a',\, b'$})
+p^{+\,+}(\mbox{\boldmath $a',r$})
+p^{+\,-}(\mbox{\boldmath $a',r$}) \\  \nonumber
&&+p^{-\,+}(\mbox{\boldmath $a',r$}) 
+p^{-\,-}(\mbox{\boldmath $a',r$}) 
+p^{+\,+}(\mbox{\boldmath $r,b'$})
+p^{+\,-}(\mbox{\boldmath $r,b'$})
+p^{-\, +}(\mbox{\boldmath $r,b'$})\\  
&&+p^{-\, -}(\mbox{\boldmath $r,b'$}) \bigg ] \, \bigg / \,
\left[p^{+\,+}(\mbox{\boldmath $r,r$}) 
+p^{+\,-}(\mbox{\boldmath $r,r$})
+p^{-\,+}(\mbox{\boldmath $r,r$})
+p^{-\,-}(\mbox{\boldmath $r,r$}) \right ] \geq -1.
\end{eqnarray}
or
\begin{eqnarray} {\nonumber}
&&\bigg[E(\mbox{\boldmath $a,\, b$})
+E(\mbox{\boldmath $b',\, a$})   
+E(\mbox{\boldmath $b,\, a'$}) 
-2p^{+ +}(\mbox{\boldmath $a',\, b'$})
-2p^{- \,-}(\mbox{\boldmath $a',\, b'$})
+p^{+\,+}(\mbox{\boldmath $a',r$}) \\  \nonumber
&&+p^{+\,-}(\mbox{\boldmath $a',r$})
+p^{-\,+}(\mbox{\boldmath $a',r$}) 
+p^{-\,-}(\mbox{\boldmath $a',r$}) 
+p^{+\,+}(\mbox{\boldmath $r,b'$})
+p^{+\,-}(\mbox{\boldmath $r,b'$})
+p^{-\, +}(\mbox{\boldmath $r,b'$})\\  
&&+p^{-\, -}(\mbox{\boldmath $r,b'$}) \bigg ] \, \bigg / \,
\left[p^{+\,+}(\mbox{\boldmath $r,r$}) 
+p^{+\,-}(\mbox{\boldmath $r,r$})
+p^{-\,+}(\mbox{\boldmath $r,r$})
+p^{-\,-}(\mbox{\boldmath $r,r$}) \right ] \geq -1.
\end{eqnarray}

\noindent Note that in inequality $(40)$ [or $(41)$],
the number of emissions $N$ from the source
is eliminated from the ratio.
Using $(32)$, it can easily be seen that
quantum mechanics violates
inequality $(40)$ [or $(41)$]
in case of real experiments where the solid angle covered 
by the aperture of the apparatus, $\Omega$, is  much less than
$4 \pi$. In particular, the magnitude of violation is maximized if the
following orientations are chosen
$\left (\mbox{\boldmath $a , \, b$} \right)=
 \left( \mbox{\boldmath $b' , \, a$} \right)=
 \left( \mbox{\boldmath $b , \, a'$} \right)= 120^\circ $
and
$ \left( \mbox {\boldmath $a', \, b'$} \right)=
\left ( \mbox {\boldmath $a', \, r$} \right) =
\left ( \mbox {\boldmath $r , \, b'$} \right) = 0^\circ $.

Inequality $(41)$
may be considerably simplified if we invoke some
of the symmetries that are exhibited in atomic-cascade photon
experiments. For a pair of photons in 
cascade from state $J=1$ to $J=0$, the
quantum mechanical detection probabilities $p^{\pm\, \pm}_{QM}$ and
expected value $E_{QM}$ exhibit the following
symmetry
\begin{eqnarray} 
p^{\pm\, \pm}_{QM} \,\left (\mbox{\boldmath $a,b$} \right)=
p^{\pm\, \pm}_{QM}
\,\left ( \mid\mbox{\boldmath $a-b$} \mid\right), \qquad
E_{QM} \,\left (\mbox{\boldmath $a,b$} \right)=
E_{QM} \,\left ( \mid\mbox{\boldmath $a-b$} \mid\right).
\end{eqnarray}
We assume that the 
local theories also exhibit the same symmetry
\begin{eqnarray}
p^{\pm\, \pm} \,\left (\mbox{\boldmath $a,b$} \right)=
p^{\pm\, \pm} \,\left ( \mid\mbox{\boldmath $a-b$} \mid\right), \qquad
E \,\left (\mbox{\boldmath $a,b$} \right)=
E \,\left ( \mid\mbox{\boldmath $a-b$} \mid\right).
\end{eqnarray}
Note that there is no harm in assuming Eqs.
$(43)$ since they are subject to experimental test (CHSH
\cite {4}, FC \cite{5}, and CH \cite {6}
made the same assumptions).
Using the above symmetry, inequality $(41)$ may be written as
\begin{eqnarray} \nonumber
&&\bigg [E \,\left (\mid \mbox{\boldmath $a-b$} \mid \right)+
E \,\left (\mid \mbox{\boldmath $b-a'$} \mid \right)+
E \,\left (\mid \mbox{\boldmath $b'-a$} \mid \right)-
2p^{+\, +} \,\left (\mid\mbox{\boldmath $a'-b'$} \mid\right)
-2p^{-\, -} \,\left (\mid\mbox{\boldmath $a'-b'$} \mid\right) \\ 
\nonumber
&&+p^{+\, +} \,\left (\mid\mbox{\boldmath $a'-r$} \mid\right)+
p^{+\, -} \,\left (\mid\mbox{\boldmath $a'-r$} \mid\right)+
p^{-\, +} \,\left (\mid\mbox{\boldmath $a'-r$} \mid\right) 
+p^{-\, -} \,\left (\mid\mbox{\boldmath $a'-r$} \mid\right)\\
\nonumber
&&+p^{+\, +} \,\left (\mid\mbox{\boldmath $r-b'$} \mid\right)+
p^{+\, -} \,\left (\mid\mbox{\boldmath $r-b'$} \mid\right)+
p^{-\, +} \,\left (\mid\mbox{\boldmath $r-b'$} \mid\right)
+p^{-\, -} \,\left (\mid\mbox{\boldmath $r-b'$} \mid\right)
\bigg ] \, \bigg / \, \\ 
&&\bigg [p^{+\,+}\, (0^\circ) +
p^{+\,-}\, (0^\circ) +
p^{-\,+}\, (0^\circ) +
p^{-\,-}\, (0^\circ) 
\bigg ] \geq -1.
\end{eqnarray}
We now take $\mbox{\boldmath $a'$}$ and 
$\mbox{\boldmath $b'$}$ to be along direction
$\mbox{\boldmath $r$}$,
and we take
$\mbox{\boldmath $a$}$, 
$\mbox{\boldmath $b$}$, and $\mbox{\boldmath $a'$}$ to be
three coplanar axes, each making $120^\circ$ with the other
two,
that is we choose the 
the following orientations,
$\mid \mbox{\boldmath $a - b$} \mid=
 \mid \mbox{\boldmath $b' - a$} \mid=
 \mid \mbox{\boldmath $b - a'$} \mid= 120^\circ $
and 
$ \mid \mbox {\boldmath $a'- b'$} \mid= 
\mid \mbox {\boldmath $a'- r$} \mid= 
\mid \mbox {\boldmath $r- b'$} \mid= 0^\circ $.
Furthermore if we define $K$ as
\begin{eqnarray}
K=p^{+\,+}(0^\circ)
+p^{+\,-}(0^\circ)
+p^{-\,+}(0^\circ)
+p^{-\,-}(0^\circ)
\end{eqnarray}
then the above inequality is simplified to
\begin{eqnarray}
\frac {3E \left( 120^\circ \right)
+2p^{+\, -}\left( 0^\circ \right)
+2p^{- \,+}\left( 0^\circ \right)}{K} \geq -1.
\end{eqnarray}
Using the quantum mechanical probabilities
{\footnote[3]{
Note that in an ideal experiment the function $K=1$.
From $(20)$, it can be seen that $E \left( 120^\circ \right)=-0.5$
and $p^{+\, -}\left( 0^\circ \right)=
p^{- \,+}\left( 0^\circ \right)=0$. Hence inequality $(46)$ becomes
$-1.5 \ge -1$ which 
violates $(41)$ 
by a factor of $1.5$.}
[i.e., Eqs. $(38)$],
inequality $(46)$ becomes
$-1.5 \geq -1$ in the case of real experiments
(here for simplicity, we have assumed
$F \left (\theta,\phi \right ) =1$; this is a good approximation
even in the case of real experiments. In
actual experiments where the solid angle of detectors
$\phi$ is usually less than $\pi/6$, from $(33)$ it can be seen that
$F (\theta, \pi/6) \approx 0.99$).
Quantum mechanics therefore violates inequality  $(44)$ (or $(46)$]
by a factor of 1.5,
whereas it violates 
CH (or CHSH) inequality by
by a factor of $\sqrt 2$.
Thus the magnitude of violation of inequality $(46)$
is
approximately $20.7\%$ larger than the magnitude of violation of
the previous inequalities [4-10].
Inequality $(46)$ can be used to test locality 
considerably more simply than CH or CHSH inequality.

CH inequality may 
be written as
\begin{eqnarray}
\frac{3 p \left( \phi \right)- 
p \left( 3 \phi \right)- 
p \left( \mbox{\boldmath $a'$}, \infty \right)- 
p \left( \infty , \mbox{\boldmath $b$} \right)}{
p \left( \infty , \infty \right)} \leq 0.
\end{eqnarray}
The above inequality requires the measurements of five
detection probabilities:
\\
(1) The measurement of detection probability
with both polarizers set along the $22.5^\circ$ axis
[that is $p\left(22.5^\circ \right)$].
\\
(2) The measurement of detection probability
with both polarizers set along the $67.5^\circ$ axis
[that is $p\left(67.5^\circ \right)$].
\\
(3) The measurement of detection probability
with the first polarizer set along
$\mbox{\boldmath $a'$}$ axis and the second polarizer being
removed
[that is 
$ p\left(\mbox{\boldmath $a'$}, \infty \right)$].
\\
(4) The measurement of detection probability
with the first polarizer removed and the second polarizer set along
$\mbox{\boldmath $b$}$ axis
[that is 
$ p\left(\infty , \mbox{\boldmath $b$} \right)$].
\\
(5) The measurement of detection probability
with both polarizers removed [that is 
$ p \left( \infty , \infty \right)$].
\\
In contrast, the inequality derived in this paper
[i.e., inequality $(46)$] requires the measurements of only
two detection probabilities:
\\
(1) The measurement of detection probability
with both polarizers set along the $0^\circ$ axis
[that is $p^{\pm \, \pm}\left(0^\circ \right)$].
\\
(2) The measurement of detection probability
with both polarizers set along the $120^\circ$ axis
[that is $p^{\pm \, \pm}\left(120^\circ \right)$].

Inequality $(46)$ is also experimentally simpler than
FC inequality \cite{5}
(it should be noted that FC inequality is derived under the
assumptions that
(i) $p\left(  \mbox{\boldmath $a'$}, \infty \right)$ is
independent of $\mbox{\boldmath $a'$}$,
(ii) $p\left(  \mbox{\boldmath $b'$}, \infty \right)$ is
independent of $\mbox{\boldmath $b'$}$.
These assumptions, however, should be tested experimentally).
FC inequality may be written as
\begin{eqnarray}
\frac{p\left(  22.5^\circ \right)- 
p\left(  67.5^\circ \right)}{
p\left(  \infty , \infty \right)} \leq 0.25.
\end{eqnarray}
The above inequality requires the measurement of at least three
detection probabilities:
\\
(1) The measurement of detection probability
with both polarizers set along the $22.5^\circ$ axis
[that is $p\left(22.5^\circ \right )$].
\\
(2) The measurement of detection probability
with both polarizers set along the $67.5^\circ$ axis
[that is $p\left(67.5^\circ \right )$].
\\
(3) The measurement of detection probability
with both polarizers removed [that is 
$ p\left( \infty , \infty \right)$].
\\
In contrast inequality $(46)$ 
requires the measurements of only 
two detection probabilities.

\begin{center}
\Large  {\bf VI. Violation of Bell's inequality in
phase-momentum and in high-energy experiments}
\end{center}

It should be noted that the analysis that led to inequality $(46)$
is not limited to atomic-cascade experiments and
can easily be extended to experiments which use
phase-momentum \cite {14}, or use high energy
polarized protons or $\gamma$ photons [15-16] to test Bell's limit.
For example in the experiment by Rarity and Tapster \cite{14},
instead of inequality $(2)$ of their paper, the following
inequality (i.e., inequality $(46)$ 
using their notations) may be used to test locality:
\begin{eqnarray}
\frac {3E \left ( 120^\circ \right)
-2{\overline C}_{a_3 \,b_4}\left( 0^\circ \right)
-2{\overline C}_{a_4 \,b_3}\left( 0^\circ \right)}
{K} \geq -1,
\end{eqnarray}
where ${\overline C} _{a_i \,b_j} \left ( \phi_a \, , \phi_b \right)$
$(i=3,4;j=3,4)$ is the counting rate between detectors $D_{ai}$ and
$D_{bj}$ with phase angles being 
set to $\phi_a \, , \phi_b$ (See Fig. 1 of \cite {14}).
The following set of orientations
$\mbox{\boldmath $(\phi_a,\, \phi_b)$}=
\mbox{\boldmath $(\phi_{b'},\, \phi_a)$}=
\mbox{\boldmath $(\phi_b,\, \phi_{a'})$}=120^\circ $, and
$\mbox{\boldmath $(\phi_{a'},\, \phi_{b'})$}=0^\circ $
leads to the largest violation.
Similarly, in high-energy experiments and
spin correlation proton-proton scattering experiments \cite {16},
inequality $(46)$ can be used to test locality.

\pagebreak
\begin{center}
\Large  {\bf VII. Conclusion}
\end{center}

In summary, we have derived
a correlation inequality [inequality $(19)$] which can be used to test
locality.
In case of ideal
experiments, this inequality is equivalent to Bell's
original inequality of $1965$ \cite{1}
or CHSH inequality \cite{4}. However, in case of real experiments where
polarizers and detectors
are not ideal, inequality $(19)$ is a new and distinct inequality and
is not equivalent to any of
previous inequalities.

We have also demonstrated that
the conjunction of Einstein's locality
[Eq. $(3)$] with a supplementary assumption [inequality $(34)$]
leads to validity of inequality $(46)$
that is sometimes grossly violated
by quantum mechanics.
Inequality $(46)$, which may be called 
{\em strong} inequality \cite{17},
defines an experiment which can actually
be performed with present technology and which does not require
the number of emissions $N$. 
Inequality $(46)$
requires the measurements of
only two detection probabilities
(at polarizer angles $0^\circ$ and $120^\circ$),
whereas CH or CHSH inequality 
requires the measurements of five detection probabilities.
This result can be of significance for the experimental
test of locality where the time during which the source emits
particles is usually very limited and it is
highly desirable to perform the least number of measurements.
Furthermore, QM violates this inequality by a factor of $1.5$, 
whereas it violates CHSH, CH or FC inequality by a factor of $\sqrt 2$.
The larger magnitude of violation
can be very useful in actual experiments.

\pagebreak


\begin{thebibliography}{99}

\bibitem{1} J. S. Bell, Physics
1 (1966) 195.

\bibitem{2} A. Einstein, B. Podolsky, and N. Rosen, Phys. Rev.
47 (1935) 777.

\bibitem{3} For an introduction to Bell's
theorem and its experimental implications,
see F. Selleri, in {\em Quantum Mechanics Versus
Local Realism}, edited by F. Selleri (Plenum Publishing
Corporation, $1988$).

\bibitem{4} J. F. Clauser, M. A. Horne, A.  Shimony, and R. A. Holt,
Phys. Rev. Lett.  23 (1969) 880.

\bibitem{5} S. J. Freedman, and J. F. Clauser,
Phys. Rev. Lett.  28 (1972) 938.

\bibitem{6} J. F. Clauser, and M. A. Horne, Phys. Rev. D. 
10 (1974) 526.

\bibitem{7} J. F. Clauser, and A. Shimony, Rep. Prog. Phys. 
{\bf 41}, (1978) 1881, see especially pp. 1891-1896.

\bibitem{8} J. S. Bell, in {\em Foundations
of Quantum Mechanics, Proceedings of
the international School of Physics `Enrico Fermi,' Course XLIX},
edited by B. d'Espagnat (Academic, New York, 1971).

\bibitem{9} A. Garuccio, and V. Rapisarda, Nuovo Cim. A.
65 269 (1981) 269.

\bibitem{10} M. Ardehali, Phys. Rev. A 47A (1993) 1633.

\bibitem{11} M. Ardehali, Phys. Rev. A {\bf 49R}, 3145 (1994).

\bibitem{12} D. Bohm, {\em Quantum Theory} (Prentice-Hall,
Englewood Cliffs, NJ, 1951), pp. 614-823.

\bibitem{13} A. Aspect, P. Grangier, and G. Roger, Phys. Rev.
Lett. 47 (1981) 460;
A. Aspect, P. Grangier, and G. Roger, Phys. Rev.
Lett. 49 (1982) 91;
A. Aspect, P. Grangier, and G. Roger, Phys. Rev.
Lett. 49 (1982) 1804;
Z. Y. Ou, X. Y. Zou, L. J. Wang, and L. Mandel, Phys. Rev.
Lett. 65 (1990) 321;
Y. H. Shih, and C. O. Alley, Phys. Rev.
Lett. 61 (1988) 2921.

\bibitem{14} J. G. Rarity and P. R. Tapster, Phys. Rev. Lett. 64
(1990) 2921.

\bibitem{15} L. R. Kasday, J. D. Ullman, and C. S. Wu, Bull. Am.
Phys. Soc. 15 (1970) 586; A. R. Wilson, J. Lowe, and D. K.
Butt, J. Phys. G 2 (1976) 613; M. Bruno, M. d'Agostino,
and C. Maroni, Nuovo Cimento 40B (1977) 142.

\bibitem{16} M. Lamehirachti
and W. Mittig, Phys. Rev. D 14
(1976) 2543.

\bibitem{17} F. Selleri, in
{\em Microphysical Reality and Quantum Formalism},
edited by A. van der Merwe, F. Selleri, and G. Tarrozi (Kluwer,
Dordrecht, $1988$).

\end{thebibliography}
\end{document}